\documentclass[apjl]{emulateapj}

\usepackage[colorlinks,citecolor=blue]{hyperref}
\def\Msun{M$_{\odot}$}
\def\oii{[O\,{\scriptsize II}]}
\shorttitle{CO(2-1) emissions in the XMMXCS J2215.9-1738 galaxy cluster}
\shortauthors{Hayashi M. et al.}

\begin{document}

\title{Evolutionary phases of gas-rich galaxies in a galaxy cluster at z=1.46}

\author{Masao Hayashi\altaffilmark{1},
  Tadayuki Kodama\altaffilmark{2,1,7},
  Kotaro Kohno\altaffilmark{3,8},
  Yuki Yamaguchi\altaffilmark{3},  
  Ken-ichi Tadaki\altaffilmark{1,9},
  Bunyo Hatsukade\altaffilmark{3},  
  Yusei Koyama\altaffilmark{4,7},
  Rhythm Shimakawa\altaffilmark{5,1},
  Yoichi Tamura\altaffilmark{6,3},
  Tomoko L. Suzuki\altaffilmark{1}
}
\affil{$^{1}$National Astronomical Observatory of Japan, Osawa,
  Mitaka, Tokyo 181-8588, Japan; \textcolor{blue}{masao.hayashi@nao.ac.jp}}
\affil{$^{2}$Astronomical Institute, Tohoku University, Aramaki,
  Aoba-ku, Sendai 980-8578, Japan}
\affil{$^{3}$Institute of Astronomy, The University of Tokyo, Osawa,
  Mitaka, Tokyo 181-0015, Japan}  
\affil{$^{4}$Subaru Telescope, National Astronomical Observatory of
  Japan, 650 North A'ohoku Place, Hilo, HI 96720, USA}
\affil{$^{5}$UCO/Lick Observatory, University of California, 1156
  High Street, Santa Cruz, CA 95064, USA}
\affil{$^{6}$Department of Physics, Nagoya University, Furo-cho,
  Chikusa-ku, Nagoya 464-8601, Japan}  
\affil{$^{7}$Department of Astronomical Science, SOKENDAI (The
  Graduate University for Advanced Studies), Mitaka, Tokyo 181-8588}
\affil{$^{8}$Research Center for the Early Universe, The University of
  Tokyo, 7-3-1 Hongo, Bunkyo, Tokyo 113-0033, Japan}
\affil{$^{9}$Max-Planck-Institut f$\ddot{\rm u}$r Extraterrestrische
  Physik, Giessenbachstrasse, D-85748 Garching, Germany} 

\begin{abstract}
We report a survey of molecular gas in galaxies in the XMMXCS
J2215.9-1738 cluster at $z=1.46$. We have detected emission lines
from 17 galaxies within a radius of $R_{200}$ from
the cluster center, in Band 3 data of the Atacama Large
Millimeter/submillimeter Array (ALMA) with a coverage of 93 -- 95 GHz
in frequency and 2.33 arcmin$^2$ in spatial direction.
The lines are all identified as CO $J$=2--1 emission lines from
cluster members at $z\sim1.46$ by their redshifts and the colors of
their optical and near-infrared (NIR) counterparts. The line
luminosities reach down to $L'_{\rm CO(2-1)}=4.5\times10^{9}$ K km
s$^{-1}$ pc$^2$. 
The spatial distribution of galaxies with a detection of CO(2--1)
suggests that they disappear from the very center of the cluster.
The phase-space diagram showing relative velocity versus
cluster-centric distance indicates that the gas-rich galaxies have
entered the cluster more recently than the gas-poor star-forming
galaxies and passive galaxies located in the virialized region of this
cluster.
The results imply that the galaxies have experienced ram-pressure
stripping and/or strangulation during the course of infall towards the
cluster center and then the molecular gas in the galaxies at the
cluster center is depleted by star formation.  
\end{abstract}


%%%%%%%%%%%%%%%%%%%%%%%%%%%%%%%%%%%%%%%%%%%%%%%%%%%%%%%%%%%%%%%%%%
%%%%%%%%%%%%%%%%%%%%%%%%%%%%%%%%%%%%%%%%%%%%%%%%%%%%%%%%%%%%%%%%%%

\section{Introduction}
\label{sec:introduction}

Identifying physical mechanisms that cause quenching of star formation
in galaxies is one of the critical outstanding issues regarding galaxy
formation and evolution. Given that galaxy clusters in the local
Universe are dominated by quiescent early-type galaxies
\citep[e.g.,][]{Dressler1997, Peng2010, Scoville2013}, the environment
where galaxies reside must play a vital role in quenching their
star-formation activities, which eventually turns them into passively
evolving elliptical or S0 galaxies. Although various processes that
can be responsible for the environmental quenching are considered, for
example, major/minor merger, ram-pressure stripping, and
strangulation, the central process that governs the star-formation
activity and quenching in cluster galaxies has yet to be identified. 

Gas content of galaxies is a fundamental quantity that is directly
related to the star-formation activities in galaxies. The
gas-regulated models predict that the amount of molecular gas that
galaxies contain can govern the scaling relation such as the mass-star
formation rate (SFR) main sequence and mass-metallicity
relation \citep{Lilly2013,Zahid2014a}. Indeed, the gas fraction is
known to be larger in galaxies at higher redshifts upto $z\ga2$ and
the SFR of a galaxy with a given mass also becomes larger in
proportion to the gas fraction as the redshift
increases \citep[e.g.,][]{Tacconi2010,Geach2011,Saintonge2013}.
Since star-forming galaxies in high-$z$ galaxy clusters should grow to
massive early-type galaxies in the local Universe, the cluster
galaxies' gas content allows us to better understand the quenching
mechanisms of galaxies in dense environments.
However, most of the observations of molecular gas at high redshifts
have been limited to the galaxies in general fields
\citep[e.g.,][]{Carilli2013,Walter2014,Genzel2015,Silverman2015,Decarli2016,Seko2016,Tacconi2017}. 
Although several studies have surveyed molecular gas in galaxy
(proto-)clusters at $z\approx$ 1--3, CO emissions are detected from at
most a few member galaxies in each
cluster \citep{Aravena2012,Casasola2013,Ivison2013,Tadaki2014b,Dannerbauer2017}.  

XMMXCS J2215.9-1738 galaxy cluster at $z=1.457$
\citep[22$^{\rm h}$15$^{\rm m}$58$^{\rm s}$.5, -17$^\circ$38$'$02.5$''$;][]{Stanford2006} 
is one of the best-studied high-$z$ galaxy clusters.
\citet{Hayashi2010,Hayashi2014} conduct a deep narrowband imaging with
Suprime-Cam/Subaru targeting \oii\ emission lines from galaxies in
this cluster. The observation with Suprime-Cam/Subaru has identified
many star-forming galaxies in the core region
\citep[see also][]{Hilton2010,Ma2015}, which suggests that the massive
galaxies in the cluster core are still in their formation phase and 
are as active as those in the general field at similar redshifts. This
galaxy cluster is one of the best targets to probe the early phase of
environmental quenching processes.

This {\it Letter} focuses on detection of CO $J$=2--1
($\nu_{\rm rest}$ = 230.538 GHz) emission lines with ALMA in the
$z=1.46$ galaxy cluster and presents global properties of molecular
gas contents of the cluster galaxies. More detailed characteristics of
the individual galaxies will be discussed in forthcoming papers.
The velocity dispersion of the cluster member galaxies is $\sigma=720$
km s$^{-1}$ and the radius of the galaxy cluster is $R_{200}=0.8$
Mpc \citep{Hilton2010}. The cosmological parameters of $H_0=70$ km
s$^{-1}$ Mpc$^{-1}$, $\Omega_m=0.3$, and $\Omega_\Lambda=0.7$ are
adopted.

\section{Data}
\label{sec:data}

\begin{figure*}
  \begin{center}
    \includegraphics[width=0.85\textwidth, clip, trim=50 0 20 0]{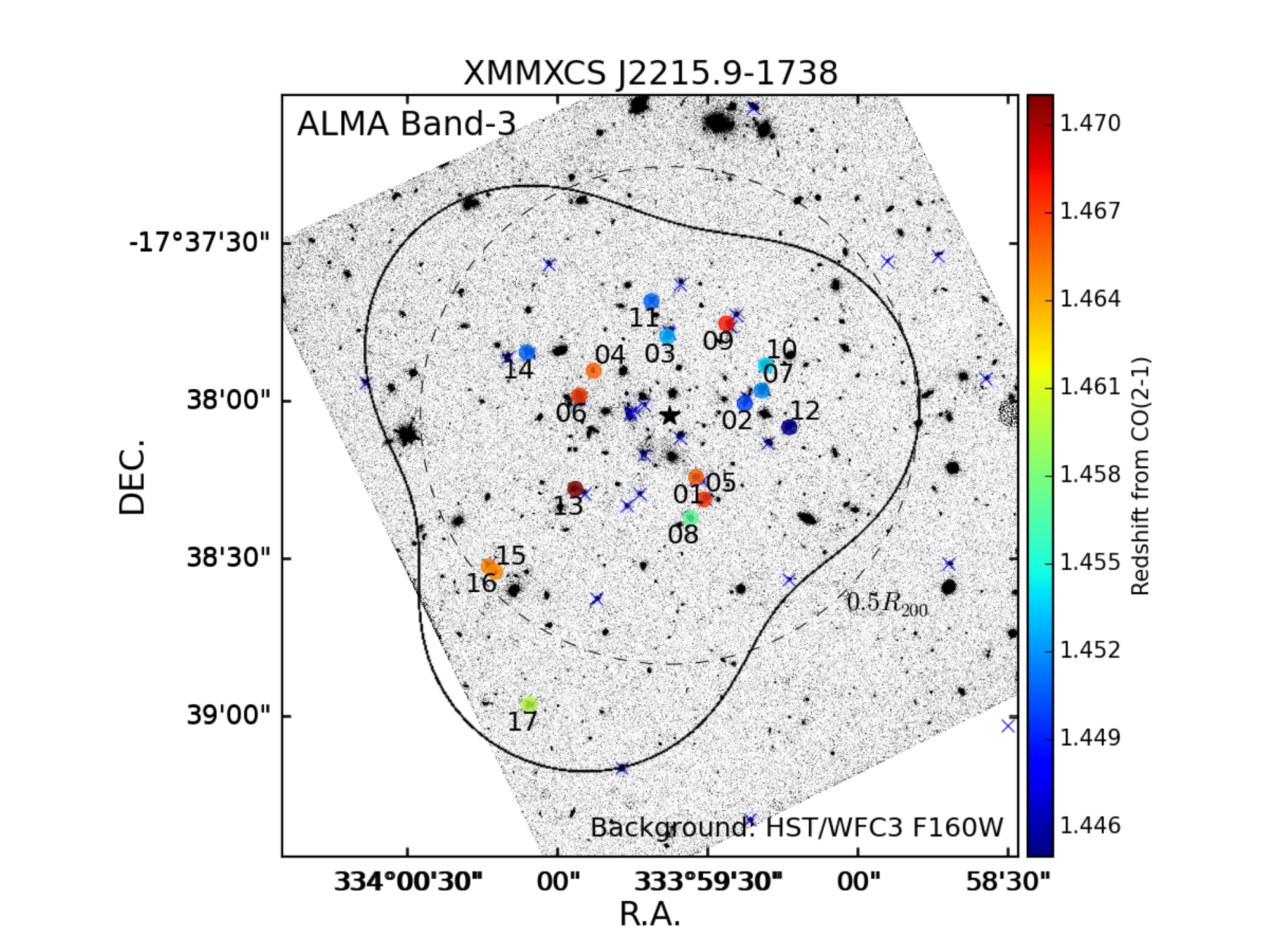}
    \caption{Spatial distribution of galaxies with detection of
      CO(2-1) line which are shown by filled circles.
      They are color-coded based on the redshifts estimated from the
      CO(2-1) lines, and the numbers next to the symbols show IDs of 
      the galaxies (see Table~\ref{table}).  
      The solid curve shows a region where the ALMA Band 3 data are
      available with a sensitivity greater than 50\%. 
      The cross symbols show the \oii\ emitters associated with this
      cluster \citep{Hayashi2014}. A star symbol shows a cluster 
      center determined with extended X-ray emission
      \citep{Stanford2006}. The background is the HST/WFC3 image in
      F160W. The dashed circle shows the cluster-centric radius of
      $0.5R_{200}$ \citep{Hilton2010}. 
      \label{fig:map}
    }
  \end{center}
\end{figure*}

Our ALMA Cycle 3 program, 2015.1.00779.S, in Band 3 was conducted in
2016 May. The spectral coverage is 93.03 -- 94.86 GHz with a spectral
resolution of 13.906 MHz ($\sim$ 12.5 km s$^{-1}$), which allows us to
capture CO(2--1) emission lines from galaxies at $z=$1.430--1.478. The
data are taken at three pointings covering a total area of 2.33 arcmin$^2$
where sensitivity is greater than 50\% (Figure~\ref{fig:map}).
Integration time is 1.04 hours per each pointing.   

Calibration of the raw data was conducted using the Common Astronomy
Software Applications \citep[CASA;][]{McMullin2007} with a standard
pipeline. The calibrated visibilities were inverted using natural
weighting to produce mosaicked 3D cubes with different velocity
resolutions (50, 100, 200, 400, and 600 km s$^{-1}$). 
The synthesized beam size is 1.79$''$ $\times$ 1.41$''$ with a position
angle of $-80$ degrees.
Typical noise levels of these cubes are 0.17, 0.12, 0.11, 0.12, and
0.12 mJy beam$^{-1}$. Note that the noise levels are measured by
taking the standard deviation of the counts in the entire cubes
including pixels with emission and thus they give upper limits of the
noise levels. Then, emission lines are extracted (see the next section
for the detailed procedure), and image deconvolution was made with a
CLEAN threshold of 5$\sigma$ of each cube, where we set CLEAN boxes of
$2'' \times 2''$, almost the same as the synthesized beam size, at the
position of each emitter.   

The optical and NIR spectroscopies already confirm 34 cluster member
galaxies within a radius of $R_{200}$ \citep{Hilton2010,Hayashi2011}.
We have created catalogs of \oii\ emission-line galaxies selected with
two narrowband filters, NB912 and NB921, \citep{Hayashi2014}. Among
the confirmed member galaxies, 20 galaxies are \oii\ emitters. We also
retrieve Hubble Space Telescope (HST) data taken with a Wide Field
Camera 3 (WFC3) (GO-13687, PI: A. Beifiori) from the HST archive. All
these data are used to search for optical and NIR counterparts of the
emission lines detected in our ALMA Band 3 data. 

\section{CO(2--1) emission lines from cluster galaxies}
\label{sec:detection}

We run {\tt Clumpfind} \citep{Williams1994} on the data cube without
the primary beam correction to search for emission lines, where we
adopt the source extraction parameters of $\Delta T = 2\sigma$ and
$T_{\rm low} = 5\sigma$. The emission line search is performed in the
cubes with different velocity resolutions of 50, 100, 200, 400, and
600 km s$^{-1}$.
We find 8, 7, 5, 7, and 7 emission line candidates at $>$ 5$\sigma$ in
each cube, and after excluding overlaps we have detected 21 candidates
at signal-to-noise ratio (SNR) of $>5.0$ in at least one velocity
resolution. 
Note that we performed the same procedure in the CLEANed cubes, but we
find the number of line candidates detected does not change. This is
because the impact of side lobes are almost negligible thanks to the
good $uv$ coverage of the ALMA data taken with 38 -- 42 antennas.   

In order to check the reliability of the extracted emission line
candidates, we perform the line search in the inverted data cubes in
the same manner by counting negative detections. We find 0, 0, 0,
2, and 3 negative detections in the data cubes with 50, 100, 200, 400,
and 600 km s$^{-1}$ resolution, respectively. This implies that there
can be about five false detections at $>5\sigma$ in our line search.

\begin{figure*}
  \begin{center}
    \includegraphics[width=0.9\textwidth]{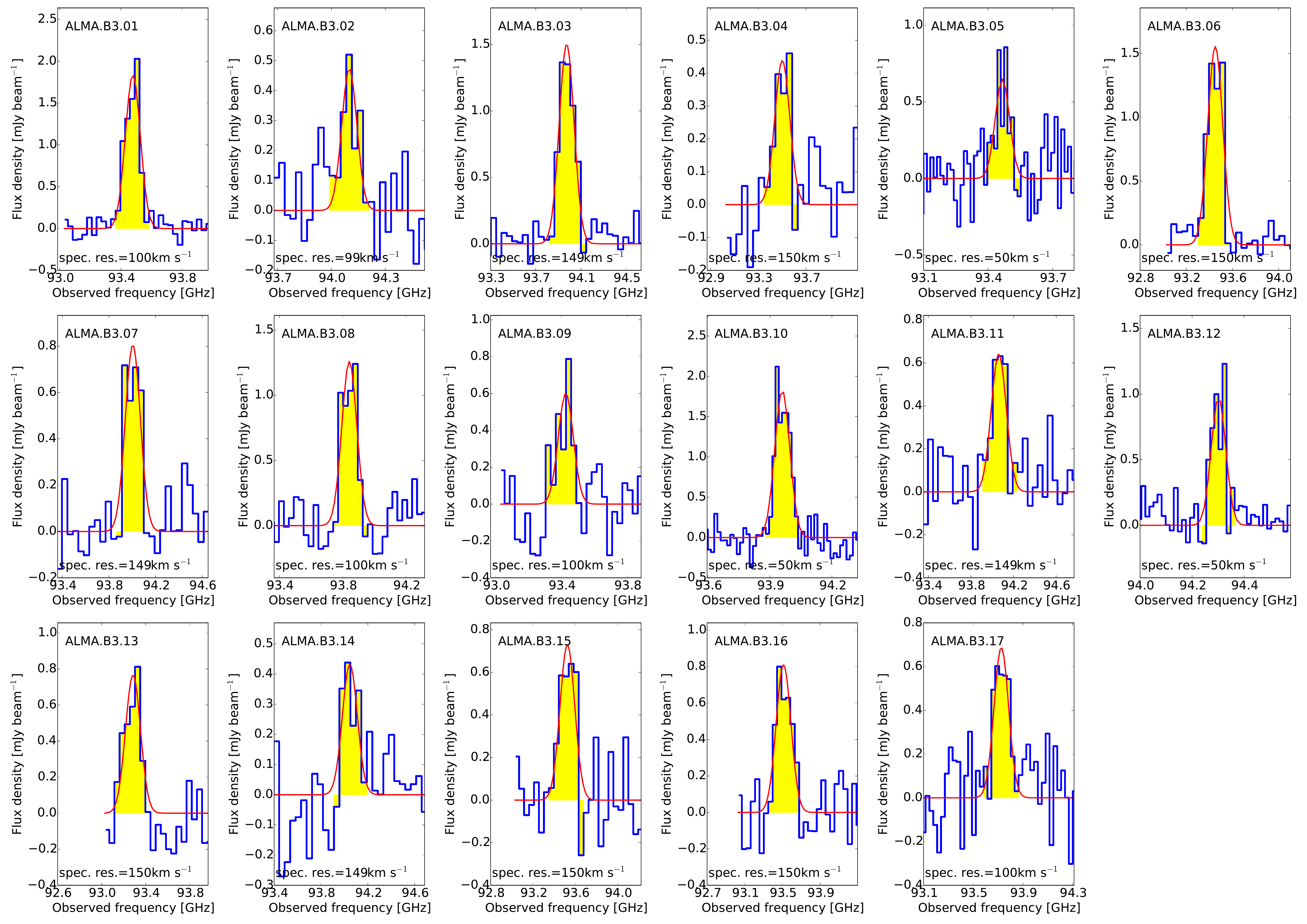}  
  \end{center}
  \caption{Spectra of the CO(2--1) emission lines, where the primary
    beam attenuation is corrected. The blue line in each panel
    shows a binned spectrum, where the spectral resolution of the binned
    spectrum is shown in each panel. The red curve shows the best fit
    Gaussian for the emission line highlighted in yellow. The
    redshift, line width, and luminosity are shown in
    Table~\ref{table}. \label{fig:spectrum}}  
\end{figure*}

To remove the possible false detections, we cross-match the
coordinates of the detections in the ALMA data with those of the
objects in the optical and NIR data catalogs by \citet{Hayashi2014}.
We use a search radius of 1 arcsec for the object matching.
We find that 17 line candidates have counterparts in the optical--NIR
data: 11 have counterparts of \oii\ emitters, and five have
counterparts of sBzK galaxies \citep{Daddi2004}. 
For the remaining one line, \#15, the counterpart is not an
\oii\ emitter, and also it does not have colors that meet the sBzK
color criteria. However, \#15 seems to be a pair of galaxies with
\#16, judging from the small spatial separation of $\sim$0.9 arcsec
and the same central frequency of the lines 
(Figures~\ref{fig:spectrum} -- \ref{fig:intensity}).
Four of the detections have no counterparts in the optical and NIR
data, which are selected from the data with a 600 km s$^{-1}$
resolution only. In this {\it Letter}, we remove these emission line
candidates without optical/NIR counterpart from the list of emission
lines detected. Note that this is consistent with the rate of false
detection we estimate.

The list of the 17 emission lines is given in Table~\ref{table}, and
their spectra are shown in Figure~\ref{fig:spectrum}.
All the detected emission lines have a single Gaussian profile and
none has a double peaked profile (Figure~\ref{fig:spectrum}). We
measure the frequency at the peak of an emission line, its width and
the peak line flux by fitting a Gaussian kernel to the spectrum. The
width of the lines ranges from 207 to 593 km s$^{-1}$.
For six of the 17 emission lines, the redshifts measured by the
optical and NIR spectroscopies are in good agreement with the
redshifts measured from the ALMA data by assuming that they are
CO(2--1) lines. Among the emission lines without any spectroscopic
redshifts, we can still estimate redshifts for five lines by combining
the narrowband imaging data of the adjacent two filters, NB912 and
NB921 \citep{Hayashi2014}. Most of the redshifts thus determined are
consistent with those of CO(2--1) lines. 

Based on the considerations, we regard the 17 lines we have detected
with ALMA as CO(2--1) lines from cluster member galaxies. We estimate
the CO(2--1) luminosities from the intensity map integrated in
velocity by the width of emission line (2$\times$FWHM)
following \citet{Solomon1992}. The luminosity, $L'_{\rm CO(2-1)}$,
ranges (4.5--22) $\times10^{9}$ K km $^{-1}$ pc$^2$. Properties of the
CO(2--1) lines are summarized in Table~\ref{table}, and their
intensity maps are shown in Figure~\ref{fig:intensity}. The detection
of 17 CO(2--1) emissions in 2.33 arcmin$^2$ indicates that the number
density of CO(2--1) emitters is several tens of times larger than that
expected from the CO luminosity function in the general fields
\citep{Walter2014,Decarli2016}.  

The spatial distribution of the CO emitters indicates that there is no
detection of CO emission line in the very center of this cluster,
i.e., $R < 0.14 R_{200}$ or 0.11 Mpc (Figure~\ref{fig:map}). The trend
is clearly seen in the top panel of Figure~\ref{fig:PSD} which shows
the cumulative fraction of galaxy populations as a function of
distance from the cluster center. The star-forming \oii\ emitters tend
to be more centrally concentrated than the CO emitters. Member
galaxies that are neither \oii\ nor CO emitters seem to be located
even closer to the cluster center, although their distribution can be
affected by a sampling bias in spectroscopic confirmation. This is
because \oii\ and CO emitters are surveyed in the narrowband imaging
or the cubic data all over the field of view, while the other members
are confirmed by slit spectroscopy. In addition, the redshifts of the
CO emitters show a bimodal distribution with the peaks at $z\sim$
1.452 and 1.466 as if they avoid the central redshift of the cluster
at $z=1.457$ as shown in the right panel of Figure~\ref{fig:PSD}. On
the other hand, the \oii\ emitters are distributed around the cluster
redshift. The results suggest that there is a difference in the
spatial distributions of galaxy populations between gas-rich galaxies
with detection of CO emission lines and the others.

\section{Discussion}
\label{sec:discussion}

\subsection{Phase-space diagram}
\label{sec:PSD}

The phase-space diagram is a useful tool to characterize the accretion
state of cluster member galaxies relatively free from effects due to
the 2D projected positions with respect to the cluster
center \citep{Noble2013,Noble2016,Jaffe2015,Muzzin2014}.  
If the motions of member galaxies are virialized around the cluster
center, the line-of-sight velocities have larger dispersions towards
the cluster center and lower dispersions at larger radii. Galaxies
that are accreted to the cluster recently tend to be offset from that
virialized relation and tend to show large relative velocities at any
radii.
\citet{Jaffe2015} define the ``virialized'' region as shown in gray in
Figure~\ref{fig:PSD} according to the orbits of dark matter haloes in
the potential of galaxy cluster based on a cosmological simulation
including a model of ram pressure. \citet{Noble2013,Noble2016} divide
the phase space into four areas (central, intermediate, recently
accreted, and infalling regions) with the curves of constant
(velocity) $\times$ (distance) to separate the accretion states of
cluster member galaxies. These defined areas on the phase-space
diagram are used to interpret our data.  

Figure~\ref{fig:PSD} shows that the CO emitters tend to be distributed
at the edge of the virialized region or in the region of relatively
recent accretion, while most of the \oii\ emitters without CO
detections and the other member galaxies that do not have strong \oii\
and CO emissions tend to be in the virialized region.
The five CO emitters (29\%) are in the ``central'' phase-space of
$(\Delta v/\sigma) \times (\Delta R/R_{200}) < 0.2$ defined by
\citet{Noble2016}, while the 17 \oii\ emitters and other member
galaxies (55\%) are in the same phase-space. 
Note that among the member galaxies without strong \oii\ and CO
emissions, some galaxies are red quiescent galaxies and the others
have a flux or equivalent width of \oii\ emission that is detectable
by spectroscopy but less than the detection limit of our narrowband
imaging \citep{Hilton2010,Hayashi2014}. Therefore the gas-rich
galaxies with CO detections have spent only relatively short times
within the cluster. 
They are likely to begin undergoing the influence of environmental
effects acting on galaxies during the course of infall to the
cluster. The other member galaxies with the amount of gas smaller than
the detection limit tend to have spent longer times as members of the 
cluster.     

\subsection{Implications for the evolution of cluster galaxies}
\label{sec:discussion}

Because each quenching mechanism is effectively at work in a specific
environment \citep{Treu2003}, the dependence of gas reservoir on the
locations in the phase-space diagram may have strong implications for
the physical processes involved in the evolution of the cluster
galaxies. Based on the above results, the following scenarios can be
considered.

\begin{figure*}
  \begin{center}
    \includegraphics[width=0.9\textwidth, clip, trim=200 75 150 50]{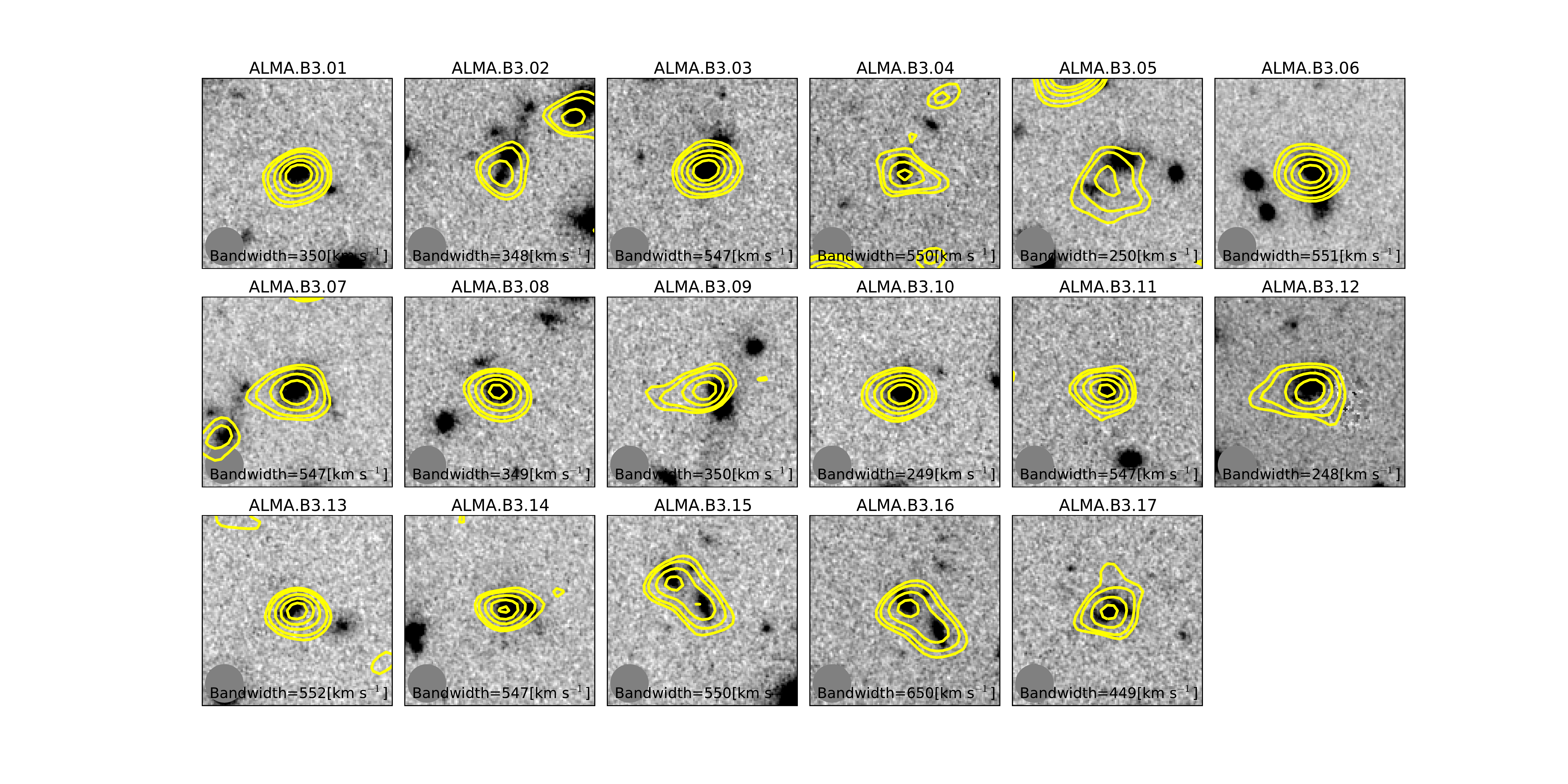}
  \end{center}
  \caption{Intensity maps of individual CO(2--1) lines overlaid on
    the HST/WFC3 F160W images of 7.5 arcsec each on a side. The
    contours show the intensity map integrated over a FWHM of each
    line around the peak frequency at 1.5, 2.0, 3.0, 4.0, and
    5.0$\sigma$ levels. The binned bandwidth is shown in the units of
    km s$^{-1}$ at the bottom of each panel. The synthesized beam size
    (spatial resolution) is shown in gray in the lower left
    corner. \label{fig:intensity}}  
\end{figure*}

\begin{figure*}
  \begin{center}  
    \includegraphics[width=1.0\textwidth, clip, trim=0 35 50 0]{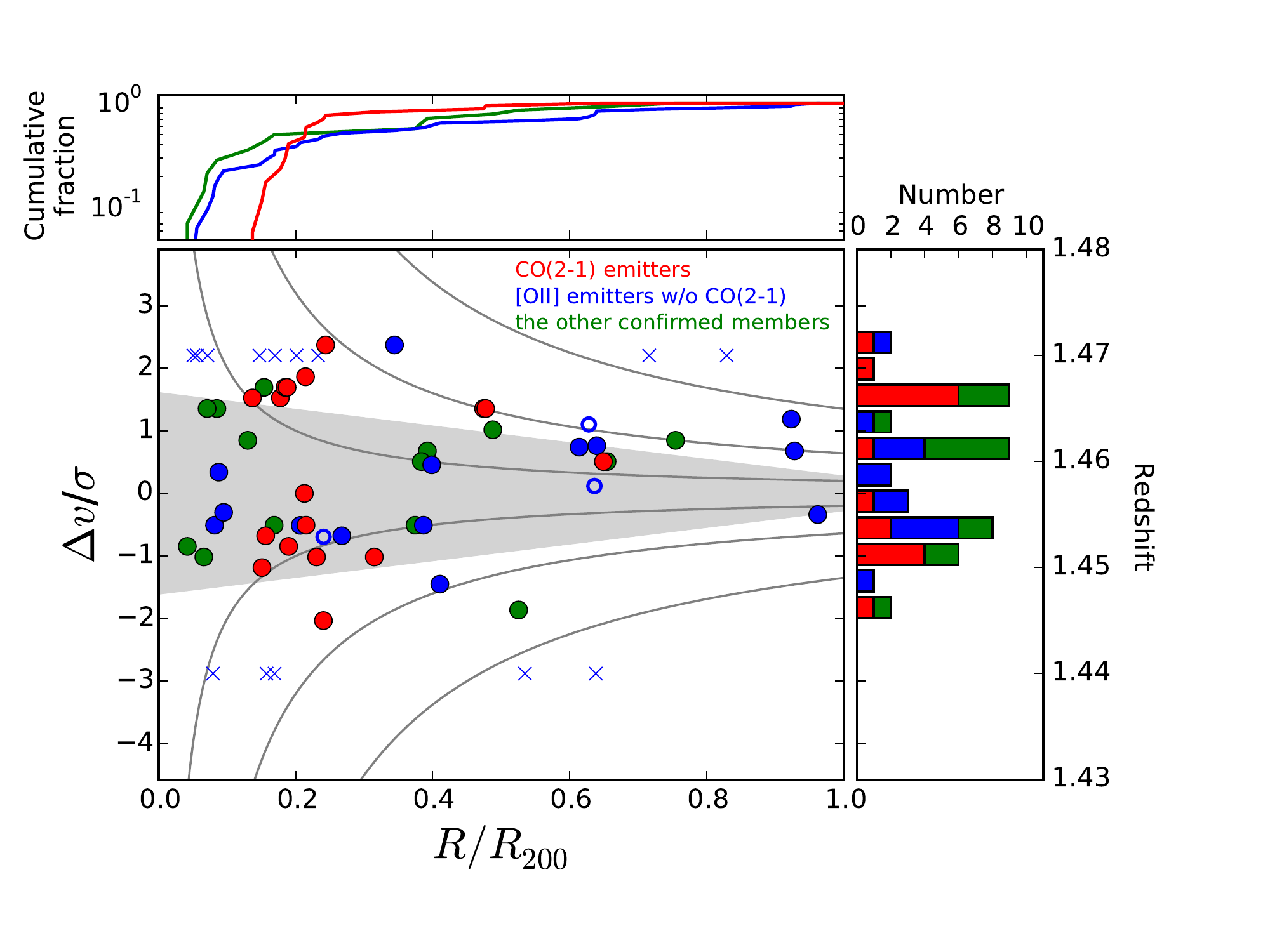}  
  \end{center}
  \caption{
    Phase-space diagram showing the relative line-of-sight velocities
    of cluster member galaxies as a function of distance from the
    cluster center. Red circles show the CO(2--1) emitters. Blue
    symbols show the \oii\ emitters without CO(2--1) detection, among
    which the filled circles represent the spectroscopically confirmed
    ones, while the open circles are those with redshifts estimated
    from the ratio of emission line fluxes of two adjacent narrowband
    filters (NB912 and NB921), and the crosses are those detected only
    either in NB912 or NB921 \citep{Hayashi2014}. We set those
    redshifts to be 1.44 (for NB912) and 1.47 (for NB921),
    respectively. Green circles show the other spectroscopically
    confirmed cluster members \citep{Hilton2010,Hayashi2014}. 
    The gray region shows a virialized area defined by
    \citet{Jaffe2015}. The gray lines show the curves of constant
    $v\times R$ values \citep{Noble2016}. 
    (right panel) Redshift histograms of spectroscopically confirmed
    galaxies. (upper panel) Cumulative fraction of the number of
    galaxies as a function of radius from the cluster center. 
    \label{fig:PSD}}
\end{figure*}

Let us first consider from star-forming galaxies in the general fields
at $z\sim1.5$ or in the outskirts of the cluster. They must have a
massive gas reservoir with the gas mass fraction of $\sim0.4$ on
average according to the previous studies \citep[e.g.,][]{Saintonge2013}.
When the galaxies are accreted onto the cluster, the first mechanisms
that can be at work include galaxy mergers or harassment. Indeed, a
merger of gas-rich galaxies (\#15 and \#16) is observed in the outer
region at $R\sim0.5R_{200}$. Then, because this cluster has hot gas
showing the extended X-ray emission \citep{Stanford2006}, during the
passage of cluster core, galaxies would suffer from ram-pressure
stripping and the gas trapped in the galaxies would be removed from
the systems. The main component of the stripped gas is HI gas
\citep[e.g.,][]{Kenney2004} and the molecular gas is relatively  
much less affected by the ram-pressure \citep[e.g.,][]{Lee2017}.
This is supported by the fact that the intensity map of all CO lines
but \#09 is in good agreement with the stellar component of the
galaxies (Figure~\ref{fig:intensity}). Therefore, CO(2--1) emission 
lines can be detected in the galaxies in the accretion region of the
phase-space diagram (Figure~\ref{fig:PSD}). The CO(2--1) luminosities
correspond to $M_{\rm H2}$=(2.0--9.4)$\times10^{10}$\Msun, under the
assumptions of $L'_{\rm CO(2-1)}/L'_{\rm CO(1-0)}=1$
\citep{Dannerbauer2009}, and $\alpha_{\rm CO}=$4.36 \citep{Tacconi2013}.
Since the star-forming member galaxies have the SFRs of
several dozens of \Msun\ yr$^{-1}$ (the median SFR = 88 \Msun\
yr$^{-1}$) \citep{Hayashi2010}, a depletion time scale is estimated to
be an order of $\sim10^{9}$ yr. This is comparable to the typical
dynamical time scale of galaxy clusters \citep{Frenk1996}. The order
estimation suggests that it is possible that a gas reservoir of a
galaxy is fully consumed by newly formed stars before it settles into
the virialized region, unless new fuel is supplied to the galaxy. If a
starvation mechanism is at work in a cluster galaxy, the HI gas is
stripped from the reservoir and the supply of fresh gas to the galaxy
is terminated. The molecular gas is consumed rapidly by remaining star
formation, and star-formation activity is eventually truncated.

Since no CO emission line is detected in the very center of the
cluster, the starvation as well as the ram-pressure stripping is
likely at work in the galaxies. By the time the galaxies settle in the
virialized region of the central region, the star-formation activity
would be fully quenched. 

To verify our scenario, the information of HI gas is also important in
addition to molecular gas. It is still impossible to observe HI gas in
and around galaxies at $z>1$, however, there were several observations
of HI gas in cluster galaxies at lower redshifts of $z\la0.2$
\citep[e.g.,][]{Verheijen2007,Lah2009,Jaffe2015,Stroe2015}.
\citet{Jaffe2015} present the result of HI gas survey in a massive
galaxy cluster at $z=0.2$ and find that almost all of the HI detected
galaxies are located in the recent infall region in the phase
space. They argue that the ram-pressure plays a key role in removing
the HI gas from the galaxies. This supports our scenario.   

Investigating the detailed properties of individual galaxies and their
dependence on the positions on the phase-space diagram can deepen our
insight into the evolution of cluster galaxies. So far, there is no
(proto-)cluster at $z>1$ with detection of CO emissions statistically
sufficient to discuss molecular gas properties of the member galaxies
on the phase-space diagram. This study, for the first time, succeed in
detecting CO(2--1) emission lines from the 17 member galaxies in the
cluster at $z=1.46$. For better understanding of the evolution of
cluster galaxies, identification of the mode of star formation,
i.e.\ starburst or secular, and the efficiency of star  formation are
key factors \citep{Daddi2010b,Genzel2010}. We will discuss these
issues in forthcoming papers.

\acknowledgments
We thank the anonymous referee for providing constructive comments and
suggestions.
MH acknowledges the financial support by JSPS Grant-in-Aid for Young
Scientists (A) Grant Number JP26707006. 
TK acknowledges the financial support in part by a Grant-in-Aid for
the Scientific Research (Nos.\ 21340045 and 24244015) by the Japanese
Ministry of Education, Culture, Sports and Science.
YT was supported by JSPS Grant-in-Aid for Scientific Research (A)
Number 15H02073.
KK was supported by JSPS Grant-in-Aid for Scientific Research (A)
Number 25247019. 
This paper makes use of the following ALMA data:
ADS/JAO.ALMA\#2015.1.00779.S. ALMA is a partnership of ESO, NSF (USA)
and NINS (Japan), together with NRC (Canada), NSC and ASIAA (Taiwan),
and KASI (Republic of Korea), in cooperation with the Republic of
Chile. The Joint ALMA Observatory is operated by ESO, AUI/NRAO and
NAOJ.

{\it Facilities:} \facility{ALMA}.

%%%%%%%%%%%%%%%%%%%%%%%%%%%%%%%%%%%%%%%%%%%%%%%%%%%%%%%%%%%%%%%%%%
%%%%%%%%%%%%%%%%%%%%%%%%%%%%%%%%%%%%%%%%%%%%%%%%%%%%%%%%%%%%%%%%%%

%% Table 1
%%%%%%%%%%%%%%%%%%%%%%%%%%%%%%%%%
\begin{deluxetable*}{cccccccc}
  \tablecaption{Properties of CO(2--1) emission lines \label{table}}
  \tablehead{
    \colhead{ID} & \colhead{R.A.} & \colhead{Dec.} &
    \colhead{redshift$^\dagger$} & \colhead{line width} &
    \colhead{$L'_{\rm CO(2-1)}$} & \colhead{SNR} &
    \colhead{counterpart} \\
    \colhead{} & \colhead{(J2000)} & \colhead{(J2000)} & \colhead{} &
    \colhead{(10$^2$ km s$^{-1}$)} & \colhead{(10$^9$ K km s$^{-1}$ pc$^2$)}
    & \colhead{} & \colhead{}
  }
  \startdata
  ALMA.B3.01 &  22 15 58.16 &  -17 38 14.5 &  1.466 &  3.7 $\pm$ 0.2 &  19.9 $\pm$ 1.1 &  18.8 &  NB921 \oii \\
  ALMA.B3.02 &  22 15 57.51 &  -17 38 00.5 &  1.450 &  3.1 $\pm$ 0.8 &  \phn 4.5 $\pm$ 0.9 &  \phn 4.7 &  sBzK \\
  ALMA.B3.03 &  22 15 58.54 &  -17 37 47.7 &  1.453 &  4.9 $\pm$ 0.3 &  21.6 $\pm$ 1.2 &  18.6 &  NB921 \oii \\
  ALMA.B3.04 &  22 15 59.52 &  -17 37 54.2 &  1.466 &  4.8 $\pm$ 1.1 &  \phn 6.2 $\pm$ 1.2 &  \phn 5.3 &  sBzK \\
  ALMA.B3.05 &  22 15 58.05 &  -17 38 18.7 &  1.467 &  2.5 $\pm$ 0.6 &  \phn 4.7 $\pm$ 0.9 &  \phn 5.1 &  NB921 \oii \\
  ALMA.B3.06 &  22 15 59.72 &  -17 37 59.0 &  1.467 &  4.9 $\pm$ 0.4 &  21.2 $\pm$ 1.2 &  18.0 &  NB921 \oii \\
  ALMA.B3.07 &  22 15 57.28 &  -17 37 58.0 &  1.452 &  4.8 $\pm$ 0.7 &  11.0 $\pm$ 1.2 &  \phn 9.5 &  sBzK \\
  ALMA.B3.08 &  22 15 58.23 &  -17 38 22.2 &  1.457 &  3.6 $\pm$ 0.4 &  12.2 $\pm$ 1.0 &  11.7 &  sBzK \\
  ALMA.B3.09 &  22 15 57.76 &  -17 37 45.2 &  1.468 &  3.5 $\pm$ 0.7 &  \phn 6.8 $\pm$ 1.0 &  \phn 6.8 &  NB912+NB921 \oii \\
  ALMA.B3.10 &  22 15 57.23 &  -17 37 53.2 &  1.454 &  2.7 $\pm$ 0.2 &  14.0 $\pm$ 0.9 &  15.4 &  NB912+NB921 \oii \\
  ALMA.B3.11 &  22 15 58.75 &  -17 37 41.0 &  1.451 &  5.3 $\pm$ 1.0 &  \phn 9.6 $\pm$ 1.2 &  \phn 8.0 &  NB912+NB921 \oii \\
  ALMA.B3.12 &  22 15 56.92 &  -17 38 05.0 &  1.445 &  2.1 $\pm$ 0.3 &  \phn 5.6 $\pm$ 0.8 &  \phn 7.3 &  NB912 \oii \\
  ALMA.B3.13 &  22 15 59.77 &  -17 38 16.7 &  1.471 &  5.2 $\pm$ 0.7 &  10.6 $\pm$ 1.2 &  \phn 8.5 &  sBzK \\
  ALMA.B3.14 &  22 16 00.41 &  -17 37 50.7 &  1.451 &  4.8 $\pm$ 1.0 &  \phn 6.2 $\pm$ 1.1 &  \phn 5.4 &  NB912 \oii \\
  ALMA.B3.15 &  22 16 00.83 &  -17 38 32.5 &  1.465 &  5.2 $\pm$ 0.9 &  10.9 $\pm$ 1.2 &  \phn 8.9 &  \\
  ALMA.B3.16 &  22 16 00.92 &  -17 38 31.5 &  1.465 &  5.9 $\pm$ 0.8 &  13.6 $\pm$ 1.3 &  10.6 &  NB921 \oii \\
  ALMA.B3.17 &  22 16 00.38 &  -17 38 57.7 &  1.460 &  4.4 $\pm$ 0.8 &  \phn 8.9 $\pm$ 1.1 &  \phn 8.2 &  NB912+NB921 \oii
  \enddata
  \tablenotetext{$\dagger$}{The redshifts are derived from CO(2-1)
    emission lines.}
\end{deluxetable*}

%% Figure 1
%%%%%%%%%%%%%%%%%%%%%%%%%%%%%%%%%
%%%%%%%%%%%%%%%%%%%%%%%%%%%%%%%%%

%% Figure 2
%%%%%%%%%%%%%%%%%%%%%%%%%%%%%%%%%
%%%%%%%%%%%%%%%%%%%%%%%%%%%%%%%%%

%% Figure 3
%%%%%%%%%%%%%%%%%%%%%%%%%%%%%%%%%
%%%%%%%%%%%%%%%%%%%%%%%%%%%%%%%%%

%% Figure 4
%%%%%%%%%%%%%%%%%%%%%%%%%%%%%%%%%
%%%%%%%%%%%%%%%%%%%%%%%%%%%%%%%%%

%%%%%%%%%%%%%%%%%%%%%%%%%%%%%%%%%%%%%%%%%%%%%%%%%%%%%%%%%%%%%%%%%%
%%%%%%%%%%%%%%%%%%%%%%%%%%%%%%%%%%%%%%%%%%%%%%%%%%%%%%%%%%%%%%%%%%

%%%%%%%%%%%%%%%%%
%%% References
%%%%%%%%%%%%%%%%%

\end{document}